\documentclass[prl,twocolumn,showpacs]{revtex4}
\usepackage{graphicx}
\usepackage{color}

\begin{document}
\tighten
\newcommand {\be}{\begin{equation}}

\newcommand {\ee}{\end{equation}}

\newcommand {\bea}{\begin{eqnarray}}

\newcommand {\eea}{\end{eqnarray}}

\def\lsim{\:\raisebox{-0.5ex}{$\stackrel{\textstyle<}{\sim}$}\:}

\def\gsim{\:\raisebox{-0.5ex}{$\stackrel{\textstyle>}{\sim}$}\:}

\def\Dtens{\mbox{\sffamily\bfseries D}}

\def\Wtens{\mbox{\sffamily\bfseries W}}

\def\Ptens{\mbox{\sffamily\bfseries P}}

\def\Otens{\mbox{\sffamily\bfseries O}}

\def\Qtens{\mbox{\sffamily\bfseries Q}}

\def\Q{\mbox{\sffamily\bfseries Q}}

\def\Ntens{\mbox{\sffamily\bfseries N}}

\def\Ctens{\mbox{\sffamily\bfseries C}}

\def\Itens{\mbox{\sffamily\bfseries I}}

\def\Atens{\mbox{\sffamily\bfseries A}}

\def\A{\mbox{\sffamily\bfseries A}}

\def\Ktens{\mbox{\sffamily\bfseries K}}

\def\Vtens{\mbox{\sffamily\bfseries V}}

\def\Gtens{\mbox{\sffamily\bfseries G}}

\def\ftens{\mbox{\sffamily\bfseries f}}

\def\vtens{\mbox{\sffamily\bfseries v}}

\def\nabbold{\mbox{\boldmath $\nabla$\unboldmath}}

\def\nabvec{\mbox{\boldmath $\nabla$}}

\def\sigtens{\mbox{\boldmath $\sigma$\unboldmath}}

\def\etatens{\mbox{\boldmath $\eta$\unboldmath}}

\def\beq{\begin{equation}}

\def\bea{\begin{eqnarray}}

\def\eeq{\end{equation}}

\def\eea{\end{eqnarray}}

\title{A Reanalysis of  the Hydrodynamic Theory of Fluid, Polar-Ordered Flocks}

\author{John Toner}
\address{Department of Physics and Institute of Theoretical Science,
University of Oregon, Eugene, OR 97403, USA}
\date{\today}
\begin{abstract}
 I  reanalyze the hydrodynamic theory of fluid, polar ordered flocks. I find new linear terms in the hydrodynamic equations which slightly modify the anisotropy, but not the scaling, of the damping of sound modes. I  also find that the nonlinearities allowed {\it in equilibrium} do not stabilize long ranged order in spatial dimensions $d=2$; in accord with the Mermin-Wagner theorem. Nonequilibrium nonlinearities  {\it do} stabilize long ranged order in $d=2$, as argued by earlier work. Some of these were missed by earlier work; it is unclear whether or not they change the scaling exponents in $d=2$.
 \end{abstract}
\maketitle

Flocking \cite{boids}  -- the coherent motion of large numbers of
self-propelled entities,  -- 
spans a wide range of length scales:  from
kilometers (herds of wildebeest) to microns (microorganisms
\cite{dictyo,rappel1}; mobile  macromolecules in living cells\cite{actin,microtub}).
It is also \cite{Vicsek}  a dynamical version of ferromagnetic ordering. A ``hydrodynamic" theory of flocking \cite{TT} shows 
that, unlike equilibrium ferromagnets \cite{MW},  
flocks {\it can}
spontaneously break a continuous
symmetry (rotation
invariance) by developing long-ranged order, (i.e., a non-zero average velocity $\left<\vec v (\vec r, t) \right>\ne \vec 0$) in spatial dimensions $d=2$, 
even with only short
ranged interactions.

The mechanism for this apparent violation of the "Mermin-Wagner" theorem \cite{MW} 
is fundamentally nonlinear.  A number of nonlinear terms in the hydrodynamic equations of motion become ``relevant", in the renormalization group (RG) sense, as the spatial dimension $d$ is lowered below 4, leading to a breakdown of linearized hydrodynamics \cite{FNS} which suppresses fluctuations enough  to stabilize long-ranged order possible in $d=2$.

In this paper, I revisit the formulation of the hydrodynamic theory of what I'll call "fluid, polar ordered" flocks,  by which I mean flocks that are spatially homogeneous, on average, and have $\left<\vec v (\vec r, t) \right>\ne \vec 0$. I find a few differences with the results of \cite{TT}. Some of these are minor: a few linear terms, that produce only minor modifications of the damping of the propagating sound modes predicted in \cite{TT}, were missed in that earlier work. 

My more important conclusions concern the scaling laws of two dimensional flocks.
It was originally argued \cite{TT} that the exponents characterizing the scaling of fluctuations in flocks that results from the breakdown of hydrodynamics could be determined exactly in $d=2$.
In this paper, I will argue that those arguments were incorrect, because they neglected certain other, equally important, symmetry-allowed nonlinearities in the hydrodynamic equations. These additional nonlinearities invalidate the earlier arguments, and render it impossible to determine the exact scaling laws in $d=2$, or, indeed, in any spatial dimension $d\le 4$.

{\it If} these new nonlinearities
should prove to be irrelevant, in the RG sense, in 
$d=2$, then the exact exponents predicted by \cite{TT} {\it would}, in fact, hold in $d=2$. At the moment, however, there is no compelling theoretical argument that they are irrelevant, though there is also none that they are not.

All of these non-linearities involve density fluctuations. Hence, in systems in which density fluctuations are suppressed, it {\it is} possible to obtain exact exponents in $d=2$. One class of such systems - flocks with birth and death - has been treated elsewhere\cite{Malthus}; others, such as incompressible systems\cite{cfl}, and systems with long-ranged interactions\cite{2DEG}, will be addressed in future work.\cite{me_future}

The new treatment  presented here correctly predicts that one naively relevant nonlinearity in the flocking hydrodynamic equations that is allowed even in equilibrium systems \cite{Marchetti} does {\it not} lead to any corrections to scaling (or, indeed, to any qualitatively new long-wavelength physics whatsoever); this means that the equilibrium systems described by such a model does {\it not} exhibit long-ranged order in $d=2$ (in accord with the Mermin-Wagner theorem \cite{MW}).

My discussion here is limited to ``ordered" flocks moving on a substrate: i.e., one  in which the flocking organisms spontaneously
pick a direction
to move together via purely short-ranged interactions that make
neighbors tend to
follow each other, but which do {\it not} pick out any a priori
preferred direction
for this motion. That is, the flocking spontaneously breaks rotation
invariance, as
equilibrium ferromagnetism does. Flocks moving without a substrate conserve momentum, and so have a very different hydrodynamics, which has been considered elsewhere\cite{mom cons}; I will not discuss these here. 
One specific realization of a flock on a substrate is the Vicsek
algorithm \cite{Vicsek}in its ordered state.

The  hydrodynamic theory describes the flock by continuous, coarse grained 
number density $\rho(\vec{r}, t)$ and velocity $\vec{v}(\vec{r}, t)$ 
fields. 
The hydrodynamic equations of motion governing these fields
 can in the long-wavelength limit can be written down purely on 
 symmetry grounds \cite{TT}, and are:
\begin{widetext}
\begin{eqnarray}
\partial_{t}
\vec{v}+\lambda_1(\vec{v}\cdot\vec{\nabla})\vec{v}+
\lambda_2(\vec{\nabla}\cdot\vec{v})\vec{v}
+\lambda_3 \vec{\nabla}(|\vec{v}|^2)
=
\alpha\vec{v}-\beta
|\vec{v}|^{2}\vec{v} -\vec{\nabla} P_1 -\vec{v} 
\left( \vec{v} \cdot \vec{\nabla}  P_2 \right) +
D_{B} \vec{\nabla}
(\vec{\nabla}
\cdot \vec{v})+ D_{T}\nabla^{2}\vec{v} +
D_{2}(\vec{v}\cdot\vec{\nabla})^{2}\vec{v}+\vec{f}
\nonumber \\
\label{EOM}
\end{eqnarray}
\end{widetext}

\begin{eqnarray}
\partial_t\rho +\nabla\cdot(\vec{v}\rho)=0
\label{conservation}
\end{eqnarray}
where all of the parameters $\lambda_i (i = 1 \to 3)$,
$\alpha$, $\beta$, $D_{B,T,2}$ and the  ``isotropic Pressure'' $P(\rho,
|\vec{v}|)$ and the  ``anisotropic Pressure''$P_2 (\rho, |\vec{v}|)$
are, in general, functions of the density $\rho$ and the magnitude
$|\vec{v}|$ of the local velocity. It is useful to Taylor expand $P_{1,2}$ and 
$P_2$ around the equilibrium density $\rho_0$:
\begin{eqnarray}
P_{1,2}=\sum_{n=1}^{\infty} \sigma_{1,2}^{(n)} ( |\vec{v}|)
(\rho-\rho_0)^n
\label{P rho}
\end{eqnarray}

Here $\beta$, $D_{B}$, $D_{2}$ and $D_{T}$ are all
positive, and
$\alpha < 0$ in the disordered phase and $\alpha>0$ in
the ordered state (in mean field theory).  

The
$\alpha$ and
$\beta$ terms simply make the local
$\vec{v}$ have a nonzero magnitude $v_0=\sqrt{{\alpha} \over
{\beta}}$ \cite{implicit}  
in the ordered phase, where
$\alpha>0$. $D_{B,T,2}$ are the diffusion constants (or
viscosities) reflecting the tendency of a localized fluctuation in the
velocities to spread out because of the coupling between
neighboring ``birds".   The $\vec{f}$ term is a random
driving force representing the noise. It is  assumed to be Gaussian with
white noise correlations:
\begin{eqnarray}
   <f_{i}(\vec{r},t)f_{j}(\vec{r'},t')>=\Delta
\delta_{ij}\delta^{d}(\vec{r}-\vec{r'})\delta(t-t')
\label{white noise}
\end{eqnarray}
   where $\Delta$ is a constant, and $i$ , $j$ denote
Cartesian components. The pressure $P$ tends, as in an equilibrium fluid, 
to maintain the local number density
$\rho(\vec{r})$ at its mean value $\rho_0$,
and $\delta \rho = \rho -
\rho_0$.   The ``anisotropic pressure'' $P_2(\rho, |\vec{v}|)$ in equation
(\ref{EOM}) is only allowed due to the non-equilibrium nature of the
flock; in an equilibrium fluid such a term is forbidden, since Pascal's
Law ensures that pressure is isotropic. In the nonequilibrium steady
state of a flock, no such constraint applies. In earlier work \cite{TT},
this term was ignored. Here I will show that this term 
changes none of the predictions of the hydrodynamic theory.

The final
equation (\ref{conservation}) is just conservation of bird number: we
don't allow our birds to reproduce or   die on the wing.
The interesting and novel results that arise when this constraint
is relaxed by allowing birth and death while the flock is moving will be
discussed elsewhere\cite{Malthus}.

The hydrodynamic model embodied in equations (\ref{EOM}), (\ref{conservation}), (\ref{P rho}),  and (\ref{white noise})  is equally valid  in both the
``disordered'' (i.e., non-moving) ($\alpha  < 0$) and 
``ferromagnetically ordered'' (i.e.,  moving) ($\alpha  > 0$) state . Here
I am interested in the ``ferromagnetically ordered'',
broken-symmetry phase which occurs for $\alpha>0$. In this state,   the velocity field can be written as:

\begin{eqnarray}
   \vec{v}=v_{0}\hat{x}_{\parallel}+\vec{\delta v}= (v_{0}+\delta v_\parallel)\hat{x}_{\parallel}+\vec{ v}_\perp~~,
\label{v fluc}
\end{eqnarray}
   where
 $v_{0}\hat{x}_{\parallel}=<\vec{v}>$ is the spontaneous average
value of
$\vec{v}$ in the ordered phase, and the fluctuations $\delta v_\parallel$ and $\vec{ v}_\perp$ of $\vec{v}$ about this mean velocity along and perpendicular to the direction of the mean velocity are assumed to be small. Indeed, I will be shortly be expanding the equation of motion (\ref{EOM}) in these quantities.
 Taking
$v_0=\sqrt{{\alpha} \over {\beta}}$ as discussed above \cite{implicit}, and taking the dot product of both sides of equation (\ref{EOM}) with $\vec{v}$ itself, I obtain:
\begin{widetext}
\begin{eqnarray}
{1\over 2}\left(\partial_{t}|\vec{v}|^2+(\lambda_1 + 2 \lambda_3)(\vec{v}\cdot\vec{\nabla})|\vec{v}|^2\right) + \lambda_2(\vec{\nabla}\cdot\vec{v})|\vec{v}|^2&= &(\alpha-\beta|\vec{v}|^{2})|\vec{v}|^{2}-\vec{v} \cdot \vec{\nabla}  P-|\vec{v}|^{2}\vec{v} \cdot \vec{\nabla}  P_2 +D_{B} \vec{v}\cdot\vec{\nabla}
(\vec{\nabla}\cdot \vec{v}) \nonumber \\&+& D_{T}\vec{v}\cdot\nabla^{2}\vec{v} +D_{2}\vec{v}\cdot\left((\vec{v}\cdot\vec{\nabla})^{2}\vec{v}\right)+\vec{v}\cdot\vec{f}
\label{v parallel elim}~.
\end{eqnarray}
\end{widetext}
In this hydrodynamic approach, 
we are interested only in fluctuations $\vec{\delta v}(\vec{r}, t)$ and $\delta \rho(\vec{r}, t)$
that vary slowly in space and time. (Indeed, the hydrodynamic equations (\ref{EOM}) and (\ref{conservation}) are only valid in this limit). Hence, terms involving space and time derivatives of 
$\vec{\delta v}(\vec{r}, t)$ and $\delta \rho(\vec{r}, t)$
are always negligible, in the hydrodynamic limit, compared to terms involving the same number of powers of fields without any time or space derivatives.

Furthermore, the fluctuations 
$\vec{\delta v}(\vec{r}, t)$ and $\delta \rho(\vec{r}, t)$ can themselves be shown to be small in the long-wavelength limit. Hence, we need only keep terms in equation (\ref{v parallel elim}) up to linear order in 
$\vec{\delta v}(\vec{r}, t)$ and $\delta \rho(\vec{r}, t)$. The 
$\vec{v}\cdot\vec{f}$ term can likewise be dropped, since it only leads to a term of order 
$\vec{v}_\perp f_\parallel$ in the $\vec{v}_\perp$ equation of motion, which is negligible (since $\vec{v}_\perp$ is small) relative to the $\vec{f}_\perp$ term already there.

These observations can be used to eliminate many of the terms in equation (\ref{v parallel elim}), and solve for the quantity
\begin{eqnarray}
U \equiv ( \alpha(\rho , |\vec{v}|)-\beta (\rho ,
|\vec{v}|)|\vec{v}|^2) &;
\label{Udef}
\end{eqnarray}
the  solution  is:
\begin{eqnarray}
U=\lambda_2 \vec{\nabla}\cdot\vec{v}+\vec{v}\cdot\vec{\nabla}P_2+{\sigma_1\over v_0}\partial_\parallel \delta \rho+{1\over 2 v_0}\left(\partial_t + \gamma_2 \partial_\parallel \right)\delta v_\parallel ,
\label{Usol}
\end{eqnarray}
where I've defined 
$\gamma_2\equiv(\lambda_1+2\lambda_3)v_0$.
Inserting this expression (\ref{Usol}) for $U$ back into equation (\ref{v parallel elim}) (where $U$ 
appears by virtue of its definition (\ref{Udef}),  I find that 
$P_2$ and 
$\lambda_2$ cancel out of the 
$\vec{v}$ equation of motion, leaving
\begin{widetext}
\begin{eqnarray}
\partial_{t}
\vec{v}+\lambda_1(\vec{v}\cdot\vec{\nabla})\vec{v}+\lambda_3 \vec{\nabla}(|\vec{v}|^2)
&=&{\sigma_1\over v_0}
\vec{v} (\partial_\parallel \delta \rho)-\vec{\nabla} P + D_{B} \vec{\nabla}
(\vec{\nabla}
\cdot \vec{v})+ D_{T}\nabla^{2}\vec{v} +
D_{2}(\vec{v}\cdot\vec{\nabla})^{2}\vec{v}\nonumber\\&+&\left[{1\over 2 v_0}\left(\partial_t + \gamma_2 \partial_\parallel \right)\delta v_\parallel
\right]\vec{v}+\vec{f}~~.
\label{EOM2}
\end{eqnarray}
\end{widetext}
This can be made into an equation of motion for $\vec{v}_\perp$ involving only $\vec{v}_\perp(\vec{r}, t)$ and $\delta \rho(\vec{r}, t)$ by projecting perpendicular to the direction of mean flock motion $\hat{x}_\parallel$, and eliminating $\delta v_\parallel$ using equation(\ref{Usol}) and 
the expansion 
\begin{eqnarray}
U\approx-\Gamma_1\left(\delta v_\parallel +{|\vec{v}_\perp|^2\over 2 v_0}\right) - \Gamma_2 \delta \rho ~~,
\label{Uexp}
\end{eqnarray}
where 
I've defined 
\begin{eqnarray}
\Gamma_1 \equiv -\left({\partial U
 \over \partial |\vec{v}|}\right)^0_{\rho} &,
&\Gamma_2 \equiv - \left({\partial U
 \over \partial \rho}\right)^0_{|\vec{v}|}~,
\label{gamma12 def}
\end{eqnarray}
with, here and hereafter , super-
or sub-scripts
$0$ denoting functions of  $\rho$ and 
$|\vec{v}|$ evaluated at $\rho = \rho_0$ and $ |\vec{v}|=v_0$.
I've also used the expansion
(\ref{v fluc}) for the velocity in terms of the fluctuations $\delta v_\parallel$ and $\vec{ v}_\perp$ to write
\begin{eqnarray}
|\vec{v}|=v_0+\delta v_\parallel +{|\vec{v}_\perp|^2\over 2 v_0}+O(\delta v_\parallel ^2, |\vec{v}_\perp|^4)~,
\label{speed}
\end{eqnarray}
and kept only terms that an RG analysis shows to be relevant in the long wavelength limit.
Inserting  (\ref{Uexp}) into (\ref{Usol})  gives:

\begin{widetext}
\begin{eqnarray}
-\Gamma_1\left(\delta v_\parallel +{|\vec{v}_\perp|^2\over 2 v_0}\right) - \Gamma_2 \delta \rho =\lambda_2 \vec{\nabla}_\perp\cdot\vec{v}_\perp+\lambda_2\partial_\parallel \delta v_\parallel+{(\sigma_{2,1} v_0^2+\sigma_{1,1})\over v_0}\partial_\parallel \delta \rho+{1\over 2 v_0}\left(\partial_t + \gamma_2 \partial_\parallel \right)\delta v_\parallel~~,
\label{v par 1}
\end{eqnarray}
\end{widetext}
where I've kept only linear terms on the right hand side of this equation, since the non-linear terms are at least of order derivatives of $|\vec{v}_\perp|^2$, and hence negligible, in the hydrodynamic limit, relative to the  $|\vec{v}_\perp|^2$ term explicitly displayed on the left-hand side.

This equation can be solved iteratively for $\delta v_\parallel$ in terms of $\vec{v}_\perp$, $\delta \rho$, and its derivatives. To lowest (zeroth) order in derivatives, $\delta v_\parallel \approx -{\Gamma_2\over \Gamma_1} \delta\rho$. Inserting this approximate expression for $\delta v_\parallel $ into equation (\ref{v par 1}) everywhere  $\delta v_\parallel$ appears on the right hand side of that equation gives $\delta v_\parallel$ to first order in derivatives:
\begin{widetext}
\begin{eqnarray}
\delta v_\parallel\approx -{\Gamma_2\over \Gamma_1}\left( \delta\rho-{1\over v_0\Gamma_1}\partial_t \delta \rho+{\lambda_4
\partial_\parallel \delta\rho\over\Gamma_2}\right)-{\lambda_2\over\Gamma_1} \vec{\nabla}_\perp\cdot\vec{v}_\perp-{|\vec{v}_\perp|^2\over 2 v_0}~,
\label{v par 2}
\end{eqnarray}
\end{widetext}
where
$\lambda_4$ is a constant related to the constants in equation (\ref{v par 1}).


 
Inserting (\ref{v fluc}), (\ref{speed}), and 
(\ref{v par 2}) into the equation of motion (\ref{EOM2}) for
$\vec{v}$, and projecting that equation perpendicular to the mean direction of flock motion $\hat{x}_\parallel$ 
gives, neglecting ``irrelevant'' terms:
\begin{widetext}
\begin{eqnarray}
\partial_{t} \vec{v}_{\perp} + \gamma\partial_{\parallel} 
\vec{v}_{\perp} &+& \lambda^0_1 \left(\vec{v}_{\perp} \cdot
\vec{\nabla}_{\perp}\right) \vec{v}_{\perp} =-g_1\delta\rho\partial_{\parallel} 
\vec{v}_{\perp}-g_2\vec{v}_{\perp}\partial_{\parallel}
\delta\rho-g_3\vec{v}_{\perp}\partial_t
\delta\rho -{c_0^2\over\rho_0}\vec{\nabla}_{\perp}
\delta\rho -g_4\vec{\nabla}_{\perp}(\delta \rho^2)\nonumber\\&+&
D^0_{B\rm{eff}}\vec{\nabla}_\perp\left(\vec{\nabla}_\perp\cdot\vec{v}_\perp\right)+
D^0_T\nabla^{2}_{\perp}\vec{v}_{\perp} +
D^0_{\parallel}\partial^{2}_{\parallel}\vec{v}_{\perp}+\nu_t\partial_t\vec{\nabla}_{\perp}\delta\rho+\nu_\parallel\partial_\parallel\vec{\nabla}_{\perp}\delta\rho+\vec{f}_{\perp} 
\label{vEOMbroken}
\end{eqnarray}
\end{widetext}
where 
$D^0_{B\rm{eff}}$,
$
D^0_{\parallel}$,  
$g_{1,2,3,4}$, 
$c_0^2$,  and 
$\nu_{t,\parallel}$ are all constants expressible in terms of the parameters in equation (\ref{v par 2}).


 
Using  (\ref{v fluc}) and (\ref{speed}) in the equation of motion (\ref{conservation}) for $\rho$ gives, again neglecting irrelevant terms:
\begin{widetext}
\begin{eqnarray}
\partial_t\delta
\rho +\rho_o\vec{\nabla}_\perp\cdot\vec{v}_\perp
+\vec{\nabla}_\perp\cdot(\vec{v}_\perp\delta\rho)+v_2
\partial_{\parallel}\delta
\rho =D_{\rho\parallel}\partial^2_\parallel\delta\rho+D_{\rho v} \partial_{\parallel}
\left(\vec{\nabla}_\perp \cdot \vec{v}_{\perp}\right)+w_1\partial_t\partial_\parallel\delta\rho
+w_2\partial_\parallel(\delta\rho^2)+{\rho_0\over2v_0}\partial_\parallel(|
\vec{v}_\perp|^2)~, \nonumber \\
\label{cons broken}
\end{eqnarray}
\end{widetext}
where 
$v_2$, 
$w_{1,2}$, and 
$D_{\rho\parallel,\rho v}$ are also  all constants expressible in terms of the parameters in equation (\ref{v par 2}). 

I will henceforth focus my attention on the uniform ordered state, in which all of the diffusion constants $
D_{\rho\parallel}$, $
D_{\rho v}$, $
D^0_{B\rm{eff}}$, $D^0_{\parallel}$, and $D_T^0$ are positive.

The linearized forms of  equations
(\ref{vEOMbroken}) and (\ref{cons broken})
differ from the  corresponding equations considered in \cite{TT} only in the $\nu_{t,\parallel}$ terms in  equation (\ref{vEOMbroken}),  and the $D_{\rho\parallel}$ and $D_{\rho v}$ terms in equation
(\ref{cons broken}). These prove \cite{me_future} to lead only to minor changes in the propagation direction dependence, but not the scaling with wavelength, of the damping of
propagating sound modes predicted in \cite{TT}; their direction-dependent speeds are unaffected.

The non-linear terms in equations (\ref{vEOMbroken}) and (\ref{cons broken}) are more significant. Several of these, specifically  all of those involving $\parallel$ and $t$ derivatives, were missed by \cite{TT}, in part because potential density dependences of various parameters were missed, and in part because of subtle mistakes in 
eliminating the fluctuations $\delta v_\parallel$ of the velocity along the mean direction of motion.

Equally noteworthy are the non-linear terms that are missing from equations (\ref{vEOMbroken}) and (\ref{cons broken}): all nonlinearities arising from the anisotropic pressure $P_2$ and the $\lambda_2$ nonlinearity. This in particular has the very important consequence of saving the Mermin-Wagner theorem. This is because  the $\lambda_2$  term is allowed even in equilibrium systems \cite{Marchetti}. The incorrect treatment in \cite{TT} suggested that this term {\it by itself} could stabilize long-range order in $d=2$. Given that this term is allowed in equilibrium, this would imply that the Mermin-Wagner theorem would fail for such an equilibrium system. The correct treatment I've done here shows that this is not the case: the  $\lambda_2$ term by itself  cannot stabilize long ranged order in $d=2$, since the non-linearities associated with it drop out of the long-wavelength description of the ordered phase.

Returning now to the non-linearities in equations
(\ref{vEOMbroken}) and (\ref{cons broken}) that were missed by \cite{TT}, I note that {\it all} of them become relevant, in the renormalization group (RG) sense\cite{Ma's book}, for spatial dimensions  $d\le d_c$, (where $d_c$ is the critical dimension below which nonlinear effects become important; here $d_c=4$), according to simple power counting. This invalidates the arguments given in \cite{TT} 
that gave exact exponents in $d=2$, since those arguments were based, inter alia, on the assertion that all of the relevant nonlinearities could, in $d=2$, be written as total $\perp$ derivatives. This assertion is clearly {\it not} true of the terms involving $\parallel$ and $t$ derivatives in equations (\ref{vEOMbroken}) and (\ref{cons broken}); hence, those terms completely invalidate the arguments leading to the exact exponents. I have been unable to come up with alternative arguments that give exact exponents in the presence of these additional terms.

Now, {\it if} these  additional nonlinearities were irrelevant in $d=2$ under a full dynamical RG, then the exact exponents of \cite{TT} would be correct in $d=2$.

There is a precedent for this (that is, for terms that appear relevant by simple power counting below some critical dimension $d_c$ actually proving to be irrelevant once "graphical corrections" -i.e., nonlinear fluctuation effects - are taken into account). One example of this is the cubic symmetry breaking interaction\cite{Aharony} in the $O(n)$ model, which is relevant by power counting at the Gaussian fixed point for $d<4$, but proves to be irrelevant, for sufficiently small $n$, at the Wilson-Fisher fixed point that actually controls the transition for $d<4$, at least for $\epsilon\equiv4-d$ sufficiently small.

Unfortunately, doing a similar $4-\epsilon$ analysis of the relevance of these new nonlinearities in the flocking problem would tell us nothing about whether or not
these terms are relevant in $d=2$, since $d=2$ is far below the critical dimension $d_c=4$ of the flocking problem.

Hence, whether or not the exact exponents predicted by \cite{TT} are correct remains an open question. They could be; numerical experiments\cite{TT, Chate1, Chate2}, and some real experiments\cite{Rao} agree with the exponents predicted by \cite{TT}, which suggests they are, but there is really no way to be certain at this point.

Not all of the predictions of \cite{TT} are problematic,
however. In particular, the claim that long ranged orientational order can exist even in $d=2$ is unaffected. We know this because the nonlinear terms clearly make positive contributions  to the velocity diffusion ``constants" $D^0_{B\rm{eff}}$ and $D^0_{T}$, and that they are 
relevant in the RG sense, which means they must change the scaling of the velocity fluctuations from that predicted by the linearized theory. We know that they are relevant by the following  proof by contradiction:  if all of the nonlinear effects were irrelevant, then simple power counting would suffice to determine their relevance. But simple power counting says that {\it all} of the nonlinearities are {\it relevant} for $d<d_c=4$, which contradicts the original assumption that they're all irrelevant.
Thus, the nonlinearites {\it must} change the scaling of the velocity fluctuations. Since the effect of the nonlinearities is to renormalize the velocity diffusion ``constants" $D^0_{B\rm{eff}}$ and $D^0_{T}$  upwards, and since this tends to reduce velocity fluctuations, the growth of velocity fluctuations with length scale must be suppressed (more precisely, its {\it scaling} must be suppressed; i.e., it must grow like a smaller power of length scale $L$)
than is predicted by the linearized version of the equations of motion (\ref{vEOMbroken}) and (\ref{cons broken}). But those linearized  equations predict \cite{TT} only 
{\it logarithmic} divergences of velocity fluctuations with length scale in $d=2$. Hence, the real fluctuations, including nonlinear effects, must be smaller than logarithmic by some power of length scale, which means they must be {\it finite } as  $L\rightarrow \infty$. This boundedness of velocity fluctuations means that long ranged order {\it is} possible in a two-dimensional flock, in contrast to equilibrium systems with continuous symmetries.

Note that  all of the troublesome nonlinearities that make it impossible to determine exact exponents in $d=2$  involve the fluctuation $\delta\rho$ of the density $\rho$. 
Therefore, if these fluctuations could somehow be ``frozen out", it would be possible to determine exact exponents in $d=2$. 

There are a number of types of flocks  in which precisely such a freezing out of density fluctuations occurs. One class of such systems - flocks with birth and death - has been treated elsewhere\cite{Malthus}; others, such as incompressible systems\cite{cfl}, and systems with long-ranged interactions\cite{2DEG}, will be addressed in future work\cite{me_future}. In all of these systems, exact scaling exponents can be found in $d=2$.

In conclusion, I have reanalyzed the hydrodynamic theory of fluid, polar ordered flocks. In addition to identifying certain new linear terms in the hydrodynamic equations for such systems, which slightly modify the anisotropy, but not the scaling, of the damping of sound modes in flocks, I have also found that certain nonlinearities that are allowed in equilibrium, and that were predicted by earlier work\cite{TT} to stabilize long ranged order in $d=2$, in fact do not. Other nonlinearities missed by earlier work could potentially change the scaling exponents from those predicted earlier\cite{TT}; whether or not they actually do so remains an open question.

I am grateful to Yu-hai Tu, Hugues Chate, Francesco Ginelli, and Cristina Marchetti for invaluable discussions, to 
the MPIPKS, Dresden, where this work was done, for their support (financial and otherwise)  and hospitality, and to the U.S.  National Science Foundation for their financial support
through awards \# EF-1137815 and 1006171.


\begin{thebibliography}{000}
\bibitem{implicit} This should be thought of as an implicit
condition on
$v_0$, since
$\alpha$ and
$\beta$ can, in general, depend on $\left| \vec{v}\right|=v_0$.
\end{thebibliography}
\end{document}